\documentclass[showpacs,superscriptaddress,twocolumn]{revtex4}
\usepackage{amsmath}
\usepackage{graphicx}

\newcommand{\nn}{\nonumber}

\begin{document}

\title{Updated predictions for graviton and photon associated production at the LHC}
\author{Jian Wang}
\address{Department of Physics
and State Key Laboratory of Nuclear Physics and Technology, Peking
University, Beijing 100871, China}

\author{Chong Sheng Li}
\email{csli@pku.edu.cn}
\address{Department of Physics and State
Key Laboratory of Nuclear Physics and Technology, Peking University,
Beijing 100871, China}
\affiliation{Center for High Energy Physics, Peking
University, Beijing, 100871, China}


\begin{abstract}
We present updated predictions on the inclusive total cross sections,
including the complete next-to-leading order  QCD corrections, for
the graviton and photon associated production process  in the large extra
dimensions model at the LHC with a center-of-mass energy of 7 and 8 TeV,
using the parameters  according to the requirements of the ATLAS and CMS Collaborations.
Moreover, we also discuss in detail the dependence on the transverse momentum cut and
uncertainties due to the choices of scales and parton distribution functions.
\end{abstract}

\maketitle

\section{Introduction}
The large extra dimensions (LED) model provides a novel framework to explain the hierarchy between
the electroweak scale $M_{\rm EW}$ ($\sim 10^3$ GeV)
and the Planck scale $M_{\rm Pl}$ ($\sim 10^{18}$ GeV) \cite{ArkaniHamed:1998rs, ArkaniHamed:1998nn, Antoniadis:1998ig}.
In this model, the space-time has $n$ extra dimensions, and
the $(4+n)$ dimensional Planck scale $M_D$ is assumed to be around the only fundamental scale $M_{\rm EW}$, while
the large Planck scale $M_{\rm Pl}$  observed in the ordinary four dimensions is viewed as an effective scale.
In this way, the hierarchy problem is reduced to the different forms of gravitational potential
in the small and large distances \cite{ArkaniHamed:1998rs}, i.e.,
\begin{eqnarray}
  V(r) &\sim& \frac{m_1 m_2}{M_D^{n+2}}\frac{1}{r^{n+1}} \quad (r\ll R), \\
  V(r) &\sim& \frac{m_1 m_2}{M_D^{n+2}R^n}\frac{1}{r} \quad (r\gg R),
\end{eqnarray}
where $R$ is the radius of extra compact spatial dimensions.
Because the Kaluza-Klein (KK) graviton, denoted as $G$, can propagate into extra dimensions,
this model predicts striking signals at hadron colliders,
such as the large missing energy associated production processes.
Among them, a photon associated with large missing energy signature is most promising
because this signature is very simple, easy to detect, and suffers from less backgrounds,
compared to any final states containing jets.
In order to address some of the most fundamental questions in extra dimension models,
the ATLAS and CMS Collaborations have searched for new phenomena in the process
with an energetic photon and large missing transverse momentum
by using the data in 2011 and set the limit on $M_{D}$ \cite{Chatrchyan:2012tea,:2012fw}.
The precise limit on $M_D$ is based on precise theoretical prediction.
The leading-order (LO) calculations and analysis of the process $pp\to \gamma G$
were performed in Ref. \cite{Giudice:1998ck}.
We presented a complete calculation of next-to-leading order (NLO) QCD corrections to this process,
and predicted inclusive and differential cross sections at the LHC in previous work \cite{Gao:2009pn}.

\section{Numerical Results and Discussion}
In this work, according to the requirement of the ATLAS and CMS Collaborations,
we present new results for the inclusive total  cross sections for graviton and photon associated production
at the LHC with a center-of-mass energy (CME) of 7 and 8 TeV, using the parameters suggested by the experimentalists
\cite{email}.
Moreover, we discuss the dependence of the total cross sections on the transverse momentum cut
and the theoretical uncertainties due to the choices of scales and parton distribution functions (PDFs), respectively.
In our numerical calculations, the  CTEQ6L1 \cite{Pumplin:2002vw} and  CTEQ6.6 PDFs  \cite{Nadolsky:2008zw}
are used for the LO and NLO results, respectively, unless specified otherwise.
We take the LED parameters $n$ and $M_D$ as input.
Except for the scale uncertainty plots, the factorization and renormalization
scales are set equal and fixed at the geometric mean of the squared transverse masses of the photon and the graviton, i.e.,
$ \mu_F^2=\mu_R^2=\mu_0^2= p_T^{\gamma}\sqrt{  p_{T,G}^2 + m^2}$,
where $p_{T,G}$ and $m$ are the transverse momentum and mass of the graviton, respectively.
We define a jet by the following requirements:
\begin{eqnarray}
p^{\rm jet}_T & > & 30\textrm{~GeV}, \quad |\eta^{\rm jet}|  <  4.5.
\end{eqnarray}
In addition, the following cuts are applied in our numerical
calculations according to the analysis of the ATLAS collaboration \cite{:2012fw}:
\begin{eqnarray}
&&p_{T}^{\gamma} > 150 \text{~GeV},\quad |\eta^{\gamma}| < 2.37,\nn\\
&&p_{T}^{\rm miss} > 150 \text{~GeV}, \quad \Delta \phi(\gamma, p_T^{\rm miss}) > 0.4, \nn\\
&&\Delta \phi({\rm jet}, p_T^{\rm miss}) > 0.4.
\end{eqnarray}
Here $\eta^{\gamma}$ is the pseudorapidity of the photon, and
$p_{T}^{\rm miss}$ is the transverse momentum of the graviton.
We also require the photon to be isolated by requiring the separation of the
photon and the radiated parton $\Delta R \equiv \sqrt{\Delta \phi^2
+ \Delta \eta^2}$  greater than 0.4.

In general, the cross section for graviton production can be expressed as \cite{Giudice:1998ck}
\begin{equation}\label{eqs:tot}
    \frac{d \sigma}{d m} = S_{n-1}\frac{\bar{M}_{\rm Pl}^2}{M_D^{2+n}}m^{n-1}\sigma_m ,
\end{equation}
where $S_{n-1}$ is the surface of a unit-radius sphere in $n$ extra dimensions, given by
\begin{equation}
    S_{n-1}=\frac{2\pi^{n/2}}{\Gamma(n/2)},
\end{equation}
and
$\bar{M}_{\rm Pl}$ is the reduced Planck mass, defined as $\bar{M}_{\rm Pl}\equiv M_{\rm Pl}/\sqrt{8\pi}$.
Here, $\sigma_m$ is the cross section for producing a graviton of mass $m$,
and the coefficient of $\sigma_m$ denotes the number density of KK graviton modes.
Since the KK graviton mass separation  is much smaller than all the other physical scales involved,
in numerical calculations, $m$ is integrated over from very low region, near zero, to the CME of the collisions
to give the total cross sections.

\begin{table}[t]
  \centering
  \begin{tabular}{c|c|c|c|c|c|c|c}
  \hline\hline
$n$ & $M_D $ [GeV]& \multicolumn{3}{|c|}{LHC 7 TeV}& \multicolumn{3}{|c}{LHC 8 TeV} \\
\hline
&&  LO [fb] & NLO [fb] & $K$ & LO [fb]  & NLO [fb] & $K$ \\
\hline
 2 & 1000 & 101.3 & 153.6 & 1.52 & 161.6 & 247.6 & 1.53 \\
 2 & 1250 & 41.5 & 62.9 & 1.52 & 66.2 & 101.4 & 1.53 \\
 2 & 1500 & 20.0 & 30.3 & 1.52 & 31.9 & 48.9 & 1.53 \\
 2 & 1750 & 10.8 & 16.4 & 1.52 & 17.2 & 26.4 & 1.53 \\
 2 & 2000 & 6.3 & 9.6 & 1.52 & 10.1 & 15.5 & 1.53 \\
 3 & 1000 & 172.0 & 222.1 & 1.29 & 311.3 & 401.7 & 1.29 \\
 3 & 1250 & 56.4 & 72.8 & 1.29 & 102.0 & 131.6 & 1.29 \\
 3 & 1500 & 22.7 & 29.2 & 1.29 & 41.0 & 52.9 & 1.29 \\
 3 & 1750 & 10.5 & 13.5 & 1.29 & 19.0 & 24.5 & 1.29 \\
 3 & 2000 & 5.4 & 6.9 & 1.29 & 9.7 & 12.6 & 1.29 \\
 4 & 1000 & 323.6 & 383.6 & 1.19 & 667.0 & 784.4 & 1.18 \\
 4 & 1250 & 84.8 & 100.6 & 1.19 & 174.9 & 205.6 & 1.18 \\
 4 & 1500 & 28.4 & 33.7 & 1.19 & 58.6 & 68.9 & 1.18 \\
 4 & 1750 & 11.3 & 13.4 & 1.19 & 23.2 & 27.3 & 1.18 \\
 4 & 2000 & 5.1 & 6.0 & 1.19 & 10.4 & 12.3 & 1.18 \\
 5 & 1000 & 652.3 & 739.4 & 1.13 & 1532.7 & 1720.1 & 1.12 \\
 5 & 1250 & 136.8 & 155.1 & 1.13 & 321.4 & 360.7 & 1.12 \\
 5 & 1500 & 38.2 & 43.3 & 1.13 & 89.7 & 100.7 & 1.12 \\
 5 & 1750 & 13.0 & 14.7 & 1.13 & 30.5 & 34.2 & 1.12 \\
 5 & 2000 & 5.1 & 5.8 & 1.13 & 12.0 & 13.4 & 1.12 \\
 6 & 1000 & 1377.7 & 1528.3 & 1.11 & 3694.6 & 4051.2 & 1.10 \\
 6 & 1250 & 231.1 & 256.4 & 1.11 & 619.8 & 679.7 & 1.10 \\
 6 & 1500 & 53.8 & 59.6 & 1.11 & 144.2 & 158.1 & 1.10 \\
 6 & 1750 & 15.7 & 17.4 & 1.11 & 42.0 & 46.1 & 1.10 \\
 6 & 2000 & 5.4 & 6.0 & 1.11 & 14.4 & 15.8 & 1.10\\
\hline\hline
\end{tabular}
  \caption{The total cross sections for the graviton and photon associated production at
  the LHC with a CME of 7 and 8 TeV.
  The corresponding $K$ factors are also shown.}
  \label{tab:kfac}
\end{table}

In Table \ref{tab:kfac}, we show the LO and NLO total cross sections
for the graviton and photon associated production at different LED parameters  $n$ and $M_D$.
We can see that for a fixed $n$, the cross section drops fast when $M_D$ increases.
This is reasonable because increasing $M_D$ means decreasing the radius of extra dimensions.
At the same time, the radius and number of extra dimensions $n$ determine the phase space that the produced graviton can propagate into; see Eq. (\ref{eqs:tot}).
Therefore, a smaller radius, with fixed $n$, leads to a smaller phase space and thus a smaller cross section.
From Table \ref{tab:kfac}, we also find that the larger is the value of $n$, the faster the cross section drops.
For example, the cross section at NLO for $n=$ 6 (2) decreases from 1528.3 (153.6) to 6.0 (9.6) ${\rm fb}$.
This is due to the facts that the graviton can propagate into $n$ extra dimensions,
and the length of each dimension is proportional to $M_D^{-1}$.
As a consequence, for a larger $n$, the effect of reducing the phase space is more obvious.

These behaviors of the cross sections result from the structure of the space-time, irrelevant of the dynamics of the scattering process.
Thus, the LO and NLO cross sections follow the same rules, which means that
the $K$ factor, defined as the ratio of the NLO cross sections to LO ones, only depends on $n$, not sensitive to $M_D$.
From Table \ref{tab:kfac}, we can see that the $K$ factor for this process is sizable.
It changes from 1.52 to 1.11 when $n$ varies from 2 to 6.
This is because that for larger $n$, the coefficient of the cross sections $\sigma_m$ in Eq. (\ref{eqs:tot}) increases faster with the increasing of
the graviton mass $m$,
which indicates that the cross sections $\sigma_m$ with large $m$ play more important roles in contributing to the total cross sections.
In order to produce a heavy KK graviton mode, the phase space of the emitted jet in real corrections is strongly suppressed,
leading to smaller NLO cross sections and thus smaller $K$ factors with the increasing of $n$.

\begin{table}
  \centering
  \begin{tabular}{c|c|c|c|c|c|c|c|c|c}
  \hline\hline
$n$ & \multicolumn{3}{|c|}{$p^{\gamma}_T>150$ GeV} & \multicolumn{3}{|c}{$p^{\gamma}_T>200$ GeV} & \multicolumn{3}{|c}{$p^{\gamma}_T>300$ GeV} \\
\hline
&  LO  & NLO  &  & LO & NLO &  & LO & NLO & \\
&  [fb] & [fb] & $K$ & [fb]  & [fb] & $K$  & [fb]  & [fb] & $K$  \\
\hline
 2 & 161.6 & 247.6 & 1.53 & 99.3 & 145.3 & 1.46 & 43.3 & 59.3 & 1.37 \\
 3 & 311.3 & 401.7 & 1.29 & 192.1 & 244.4 & 1.27 & 84.6 & 105.8 & 1.25 \\
 4 & 667.0 & 784.4 & 1.18 & 412.5 & 487.0 & 1.18 & 181.5 & 215.5 & 1.19 \\
 5 & 1532.7 & 1720.1 & 1.12 & 947.0 & 1075.3 & 1.14 & 415.6 & 480.8 & 1.16 \\
 6 & 3694.6 & 4051.2 & 1.10 & 2276.2 & 2536.2 & 1.11 & 993.3 & 1135.6 & 1.14\\
 \hline\hline
\end{tabular}
  \caption{The dependence of total cross sections at the 8 TeV LHC on the transverse momentum cut
  applied on the photon, assuming $p_T^{\rm miss}=p_T^{\gamma}$ and $M_D=1000$ GeV. The corresponding $K$ factors are also shown.}
  \label{tab:ptgamma}
\end{table}

In order to suppress the background for the signal of a photon and missing energy,
a large transverse momentum cut on the photon is usually imposed.
When searching for the signal at the LHC with a CME of 8 TeV,
experimentalists perhaps choose a harder cut on the photon transverse momentum, compared to that at the 7 TeV LHC.
To see the change of cross sections, in Table \ref{tab:ptgamma}, we show the
dependence of total cross sections at the 8 TeV LHC on the transverse momentum cut
applied on the photon, assuming $p_T^{\rm miss}=p_T^{\gamma}$ and $M_D=1000$ GeV.
When the cut becomes harder, from $p^{\gamma}_T>150$ to $p^{\gamma}_T>200$ GeV,
the NLO total cross sections decrease by about 40$\%$.
And the NLO total cross sections at $p^{\gamma}_T>300$ GeV are only about 45$\%$ of those at $p^{\gamma}_T>200$ GeV.
It is necessary to investigate how the cross sections of the corresponding backgrounds change with the increasing of the transverse momentum cut on the photon
in order to increase the ratio of the signal to backgrounds.

\begin{figure}[tbh]
  \includegraphics[width=0.78\linewidth]{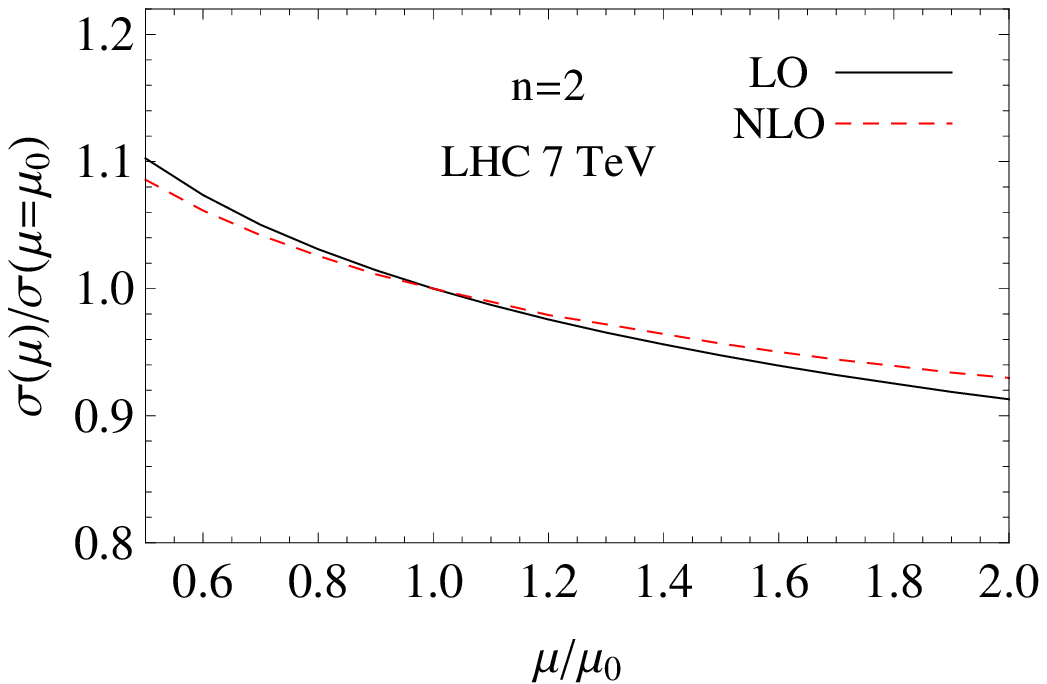}\\
  \includegraphics[width=0.78\linewidth]{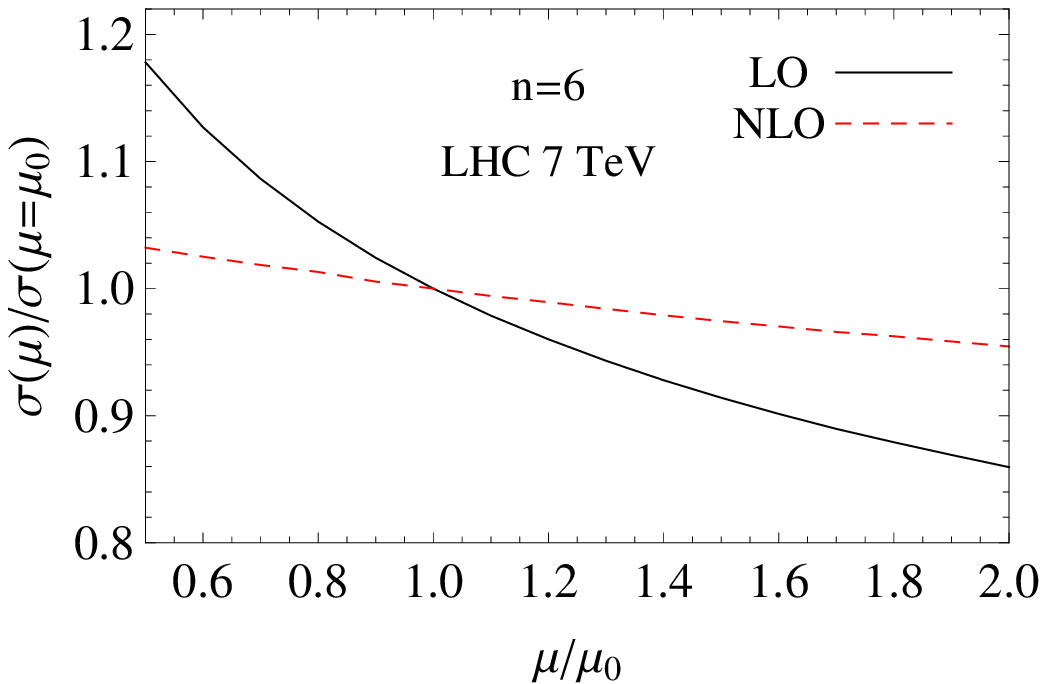}\\
  \includegraphics[width=0.78\linewidth]{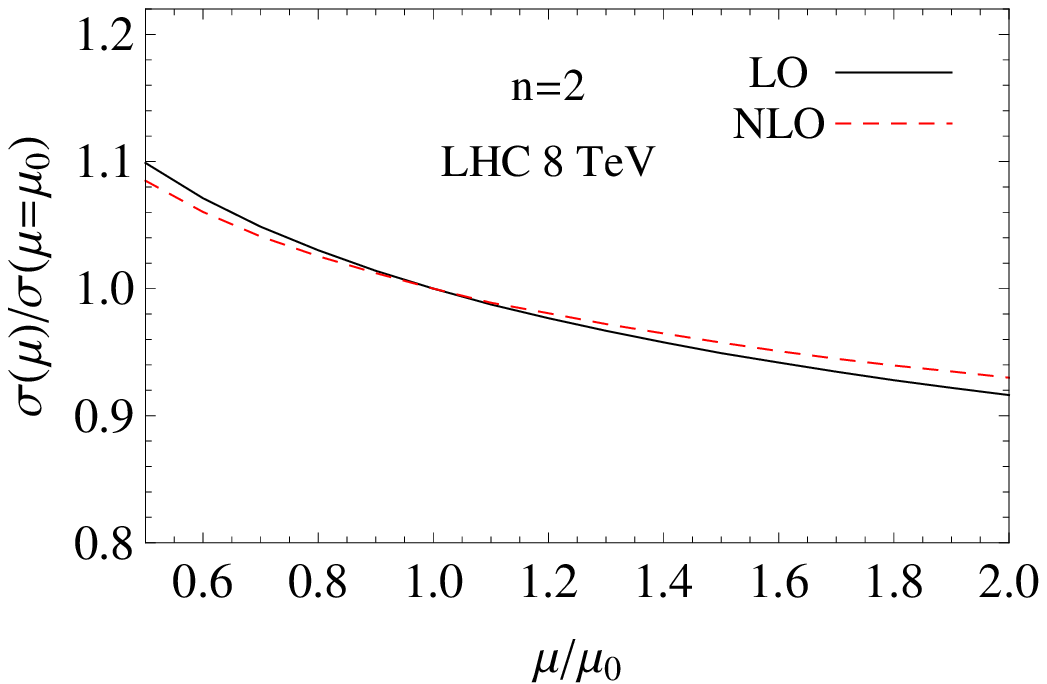}\\
  \includegraphics[width=0.78\linewidth]{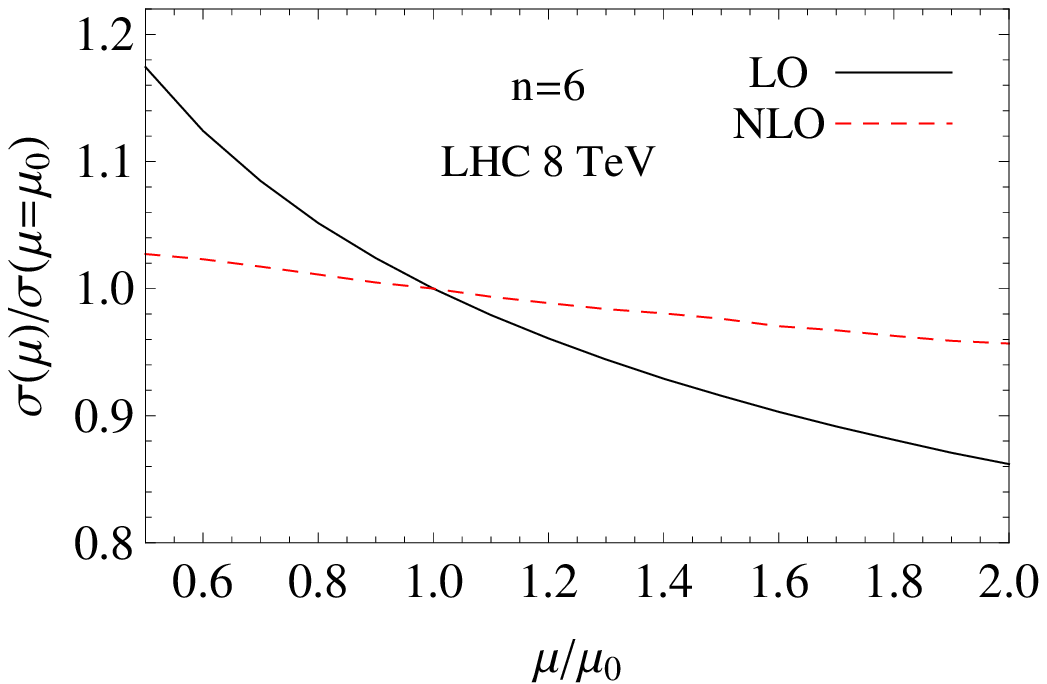}
  \caption{The scale uncertainties of the total cross sections at the LHC with a CME of 7 and 8 TeV. We have chosen $\mu_F=\mu_R=\mu$.}
  \label{fig:scl}
\end{figure}

\begin{figure}[tbh]
  \includegraphics[width=0.78\linewidth]{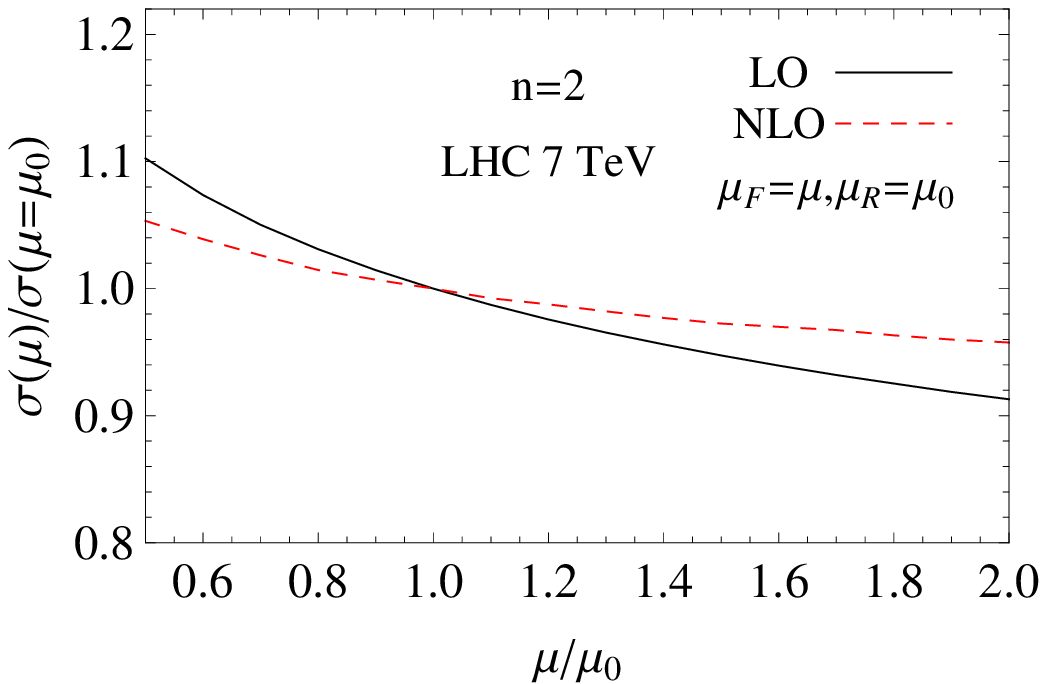}\\
  \includegraphics[width=0.78\linewidth]{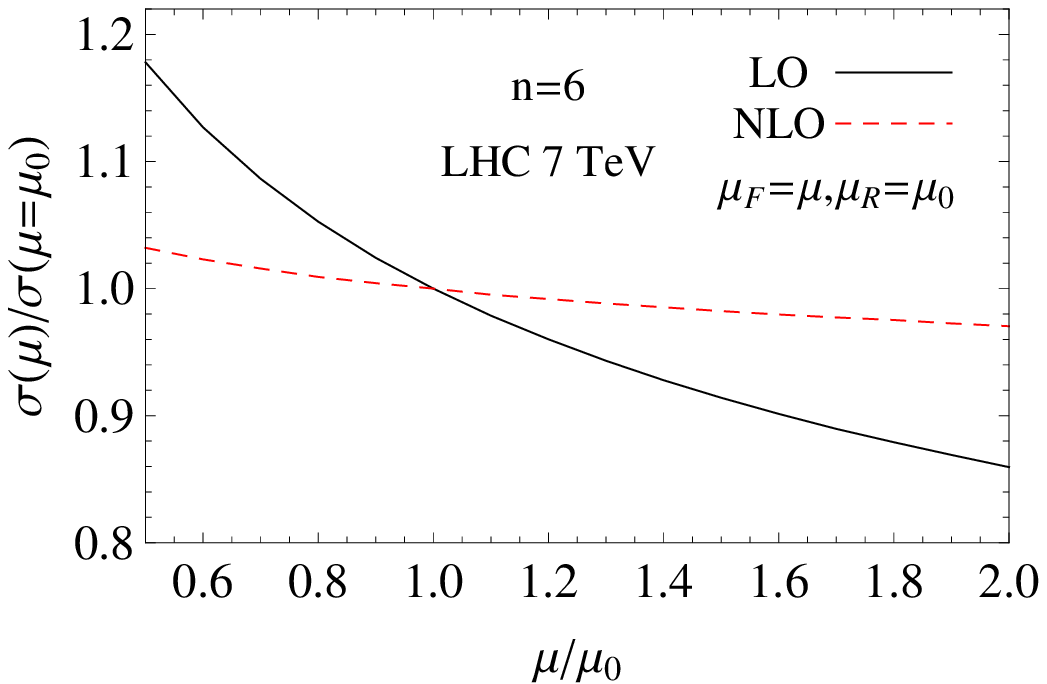}\\
  \includegraphics[width=0.78\linewidth]{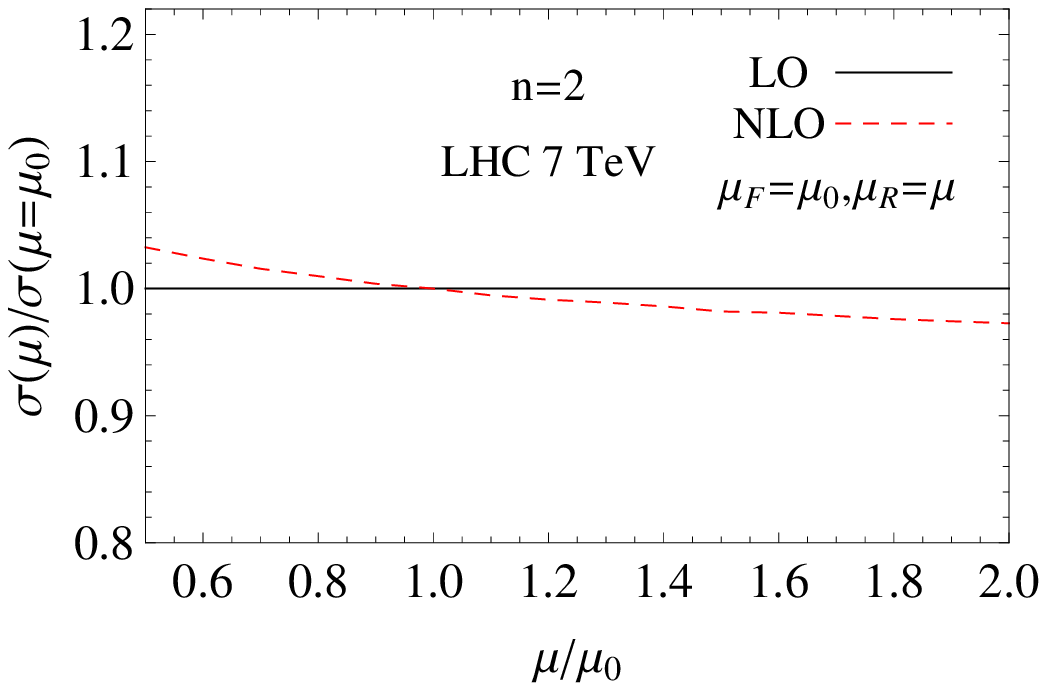}\\
  \includegraphics[width=0.78\linewidth]{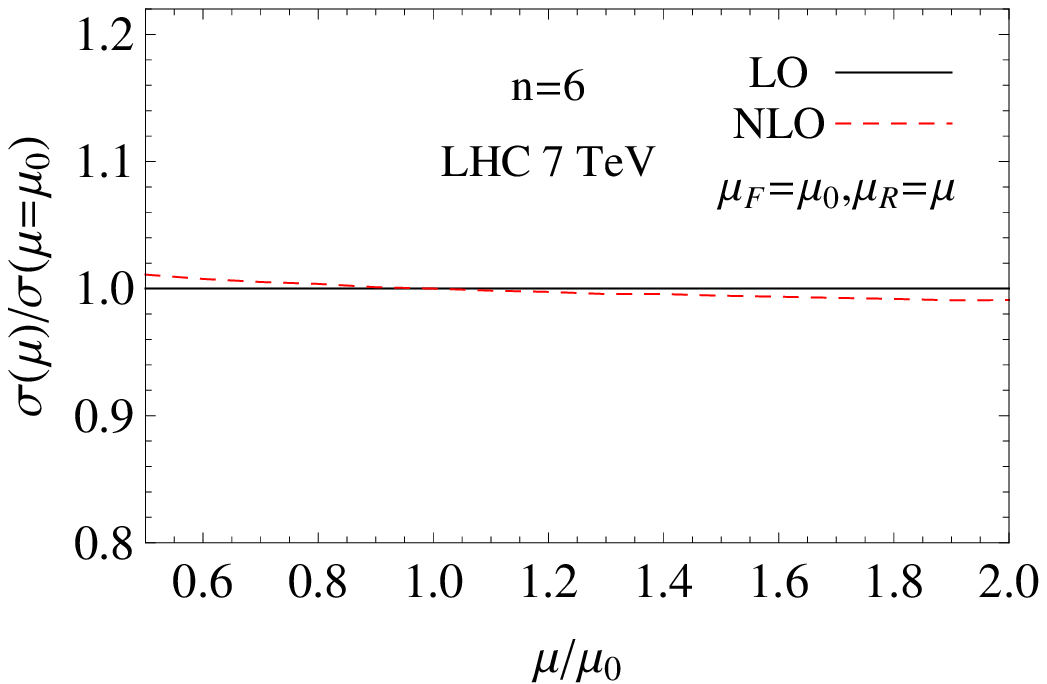}
  \caption{The scale uncertainties of the total cross sections at the LHC with a CME of 7 TeV when only one of the factorization and
  renormalization scales changes.}
  \label{fig:facrenscl}
\end{figure}

Now, we discuss the theoretical uncertainties.
First, we study the theoretical uncertainties arising from the choice of the factorization and renormalization scales.
In Fig. \ref{fig:scl}, we show the scale uncertainties of both the LO and the NLO total cross sections,
for the case of $n = 2$ and 6.
Note that these scale uncertainties are independent on $M_D$ because the cross sections are normalized by $\sigma(\mu_0)$
and thus the dependence on $M_D$ cancels.
We see that the scale uncertainties are significantly decreased by including the NLO QCD effects,
especially for the case of $n = 6$.
Explicitly, the scale uncertainties at NLO are about $\pm 8\%$ and $\pm 4\%$ for $n = 2$ and 6, respectively.
The decrease of the scale uncertainties can be more easily understood
if we study the case of changing only one of the factorization and renormalization scales, respectively, as shown in Fig. \ref{fig:facrenscl}.
The dependence of the cross section on the factorization scale is reduced due to
the inclusion of the Altarelli-Parisi splitting functions at NLO \cite{Gao:2009pn},
which accounts for scale dependence of the PDF up to $\mathcal{O}(\alpha_s)$,
and almost the same for $n$=2 and 6 at NLO.
The dependence of the cross section on the renormalization scale is absent at LO and starts at NLO,
and more obvious for $n=2$ than $n=6$ because of the larger QCD corrections for $n=2$, as shown in Table \ref{tab:kfac}.
Therefore, the decrease of the scale uncertainties for $\mu_F=\mu_R=\mu$ at NLO shown in Fig. \ref{fig:scl} is a mixing effect of the reduced factorization scale dependence and appearance of the renormalization scale dependence.
From Fig. \ref{fig:scl}, we also find that the scale uncertainties are almost the same at the LHC with a CME of 7 and 8 TeV.

\begin{figure}
  \includegraphics[width=0.78\linewidth]{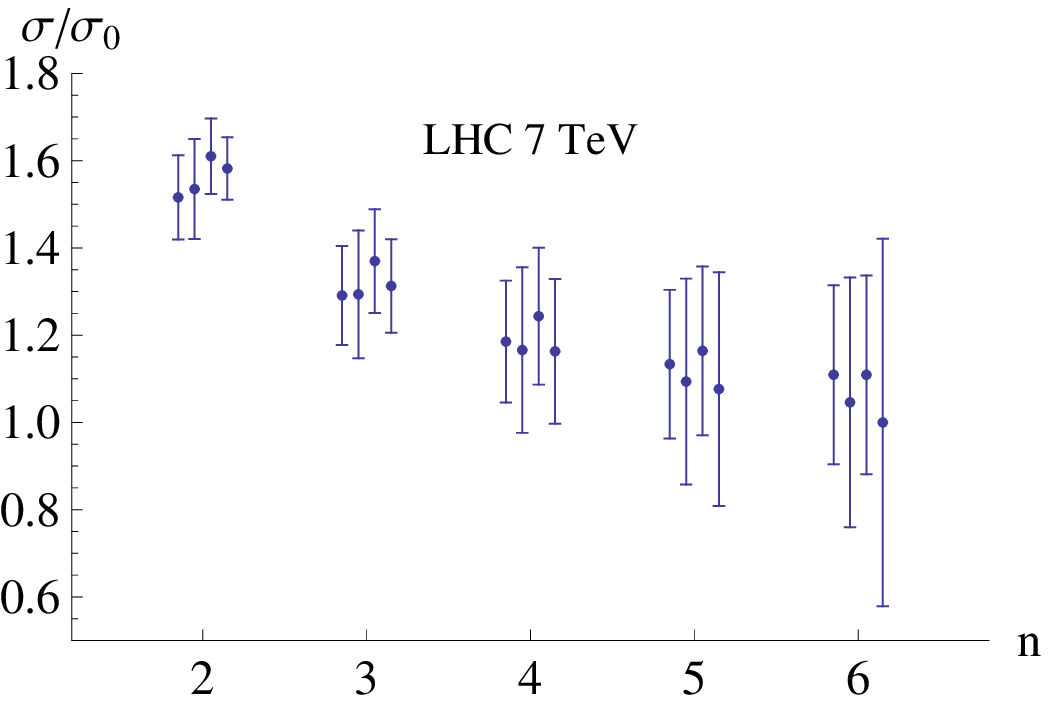}\\
  \includegraphics[width=0.78\linewidth]{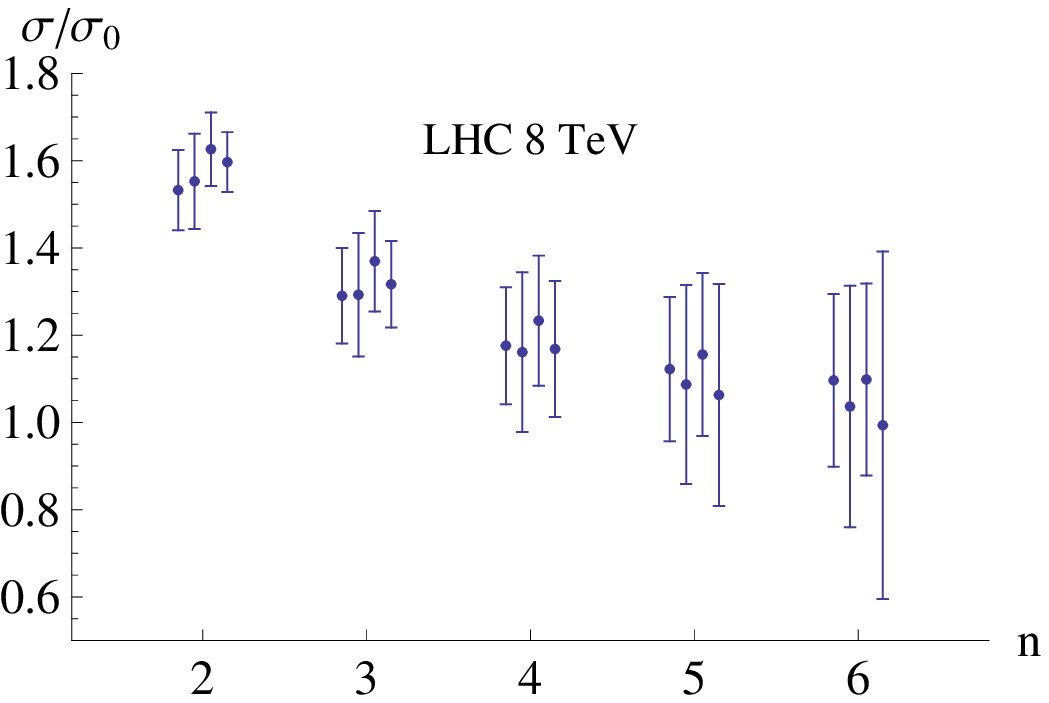}\\
  \caption{The PDF uncertainties of the NLO total cross sections.
  The lines in each group from left to right correspond to the
  CTEQ6.6, CT10, MSTW2008NLO, and NNPDF2.3NLO PDF sets.
  The PDF induced uncertainties are estimated at (or rescaled to) $90\%$ confidence-level.
  The cross section $\sigma_0$ represents the LO total cross section evaluated by using CTEQ6L1 PDF set.}
  \label{fig:PDF}
\end{figure}

In Fig. \ref{fig:PDF}, we show the PDF uncertainties of the NLO total cross sections.
We have compared the results by using the commonly used CTEQ6.6, CT10 \cite{Lai:2010vv},
MSTW2008NLO \cite{Martin:2009iq}, and NNPDF2.3NLO \cite{Nocera:2012hx} PDF sets
as recommended by the PDF4LHC Working Group in Ref. \cite{Botje:2011sn}.
Each PDF set incorporates a function to evaluate the strong coupling constant $\alpha_S$.
$\alpha_S(M_Z)$ takes the values of 0.1179, 0.1180, 0.1201 and 0.1190 for CTEQ6.6, CT10,
MSTW2008NLO, and NNPDF2.3NLO PDF sets, respectively.
The differences among  the values of $\alpha_S(M_Z)$  are less then $2\%$.
Since we have used the $\alpha_S$ function in the PDF sets to perform numerical calculations,
the PDF uncertainties actually denote the ``PDF+$\alpha_S$'' uncertainties.
From Fig. \ref{fig:PDF}, we find that the PDF uncertainties generally become larger with the increasing of $n$.
This is due to the fact that for larger $n$, the cross sections $\sigma_m$ with large $m$ play more important roles to give the total cross sections,
as explained above.
For very large $m$, the PDF reaches the region $x\to 1$,
which is not well measured and suffers from large uncertainties.
From Fig. \ref{fig:PDF}, we can also see that the predictions from various PDF sets can be very different
but are still in the range of uncertainties.

\section{Conclusion}
We have updated the predictions on
the inclusive total cross sections
for the graviton and photon associated production in the LED model at the LHC,
using the parameters  suggested by the ATLAS and CMS Collaborations.
We show the total cross sections for different LED parameters  $n$ and $M_D$
and find that the $K$ factor for this process is sizable,
changing from 1.52 to 1.11 when $n$ varies from 2 to 6, and independent on $M_D$.
We also study the dependence of total cross sections at the 8 TeV LHC on the transverse momentum cut
applied on the photon.
The NLO total cross sections decrease quickly when the cut becomes harder.
Furthermore,
we discuss the theoretical uncertainties arising from the choices of the scales and PDF sets
and find that the scale uncertainties are significantly decreased by including the NLO QCD effects,
and the predictions from various PDF sets can be very different but are still in the range of uncertainties.

\section*{Acknowledgements}
This work was supported in part by the National Natural
Science Foundation of China, under Grants No. 11021092, No. 10975004 and No. 11135003.

\bibliography{gammagv}

\end{document}